\documentclass{aa}
\usepackage{times,graphicx,url}
\usepackage{natbib}
\usepackage{bm,color}
\usepackage{latexsym}
\def\blue{\textcolor{black}}
\graphicspath{{./fig/}{./png/}}
\newcommand{\etal}{et al.}
\newcommand{\ghat}{\hat{g}}

\newcommand{\EQ}{\begin{equation}}
\newcommand{\EN}{\end{equation}}
\newcommand{\EQA}{\begin{eqnarray}}
\newcommand{\ENA}{\end{eqnarray}}

\newcommand{\Eq}[1]{Eq.~(\ref{#1})}
\newcommand{\Eqs}[2]{Eqs.~(\ref{#1}) and~(\ref{#2})}

\newcommand{\Sec}[1]{Sect.~\ref{#1}}

\newcommand{\Fig}[1]{Fig.~\ref{#1}}

\newcommand{\Figs}[2]{Figs.~\ref{#1} and \ref{#2}}

\newcommand{\Tab}[1]{Table~\ref{#1}}

\newcommand{\bra}[1]{\langle #1\rangle}

\newcommand{\meanrho}{\overline{\rho}}

{}
{}
{}

\newcommand{\meanSSSS}{\overline{\mbox{\boldmath ${\mathsf S}$}} {}}

\newcommand{\meanSSS}{\overline{\mathsf{S}}}
{}
{}
{}
{}
{}
{}
{}
{}
\newcommand{\meanAA}{\overline{\mbox{\boldmath $A$}}{}}{}
\newcommand{\meanBB}{\overline{\mbox{\boldmath $B$}}{}}{}
{}
{}
{}
{}
{}
{}
{}
{}
\newcommand{\meanJJ}{\overline{\mbox{\boldmath $J$}}{}}{}
{}
\newcommand{\meanUU}{\overline{\mbox{\boldmath $U$}}{}}{}

{}
{}
\newcommand{\meanQQ}{\overline{\mbox{\boldmath $Q$}}{}}{}

\newcommand{\meanB}{\overline{B}}

\newcommand{\meanh}{\overline{h}}

\newcommand{\meanH}{\overline{H}}
\newcommand{\meanU}{\overline{U}}

\newcommand{\meanT}{\overline{T}}

{}

{}
{}

\newcommand{\bb}{\bm{b}}
\newcommand{\BB}{\bm{B}}

\newcommand{\uu}{\mbox{\boldmath $u$} {}}
\newcommand{\UU}{\mbox{\boldmath $U$} {}}

\newcommand{\JJ}{\mbox{\boldmath $J$} {}}

\newcommand{\AAA}{\mbox{\boldmath $A$} {}}

\newcommand{\ff}{\mbox{\boldmath $f$} {}}

\newcommand{\QQ}{\mbox{\boldmath $Q$} {}}
\newcommand{\grav}{\mbox{\boldmath $g$} {}}
\newcommand{\nab}{\mbox{\boldmath $\nabla$} {}}
\newcommand{\SSSS}{\mbox{\boldmath ${\sf S}$} {}}

\newcommand{\meanDD}{{\overline{\rm D}} {}}
\newcommand{\DD}{{\rm D} {}}

\newcommand{\dd}{{\rm d} {}}

\def\la{\mathrel{\mathchoice {\vcenter{\offinterlineskip\halign{\hfil
$\displaystyle##$\hfil\cr<\cr\sim\cr}}}
{\vcenter{\offinterlineskip\halign{\hfil$\textstyle##$\hfil\cr<\cr\sim\cr}}}
{\vcenter{\offinterlineskip\halign{\hfil$\scriptstyle##$\hfil\cr<\cr\sim\cr}}}
{\vcenter{\offinterlineskip\halign{
\hfil$\scriptscriptstyle##$\hfil\cr<\cr\sim\cr}}}}}

\def\Rm{R_{\rm m}}

\def\Rey{\mbox{\rm Re}}

\def\cs{c_{\rm s}}
\def\Hpz{H_{\rm p0}}

\def\qpz{q_{\rm p0}}
\def\qp{q_{\rm p}}
\def\betap{\beta_{\rm p}}
\def\betamin{\beta_{\min}}
\def\betastar{\beta_{\star}}
\def\Peff{{\cal P}_{\rm eff}}
\def\Pmin{{\cal P}_{\rm min}}

\def\qs{q_{\rm s}}

\def\qg{q_{\rm g}}

\def\vA{v_{\rm A}}

\def\kf{k_{\rm f}}
\def\kappaf{\kappa_{\rm f}}

\def\urms{u_{\rm rms}}

\def\nut{\nu_{\rm t}}

\def\nuT{\nu_{\rm T}}

\def\etat{\eta_{\rm t}}
\def\etatz{\eta_{\rm t0}}

\def\etaT{\eta_{\rm T}}

\def\Beq{B_{\rm eq}}
\def\Beqz{B_{\rm eq0}}
\def\tautd{\tau_{\rm td}}

\def\half{{\textstyle{1\over2}}}

\def\onethird{{\textstyle{1\over3}}}

\newcommand{\Mm}{\,{\rm Mm}}

\newcommand{\yapj}[3]{ #1, {ApJ,} {#2}, #3}

\newcommand{\yapjl}[3]{ #1, {ApJ,} {#2}, #3}

\newcommand{\yan}[3]{ #1, {Astron.\ Nachr.,} {#2}, #3}

\newcommand{\yana}[3]{ #1, {A\&A,} {#2}, #3}

\newcommand{\ypf}[3]{ #1, {Phys.\ Fluids,} {#2}, #3}
\newcommand{\ypfb}[3]{ #1, {Phys.\ Fluids B,} {#2}, #3}

\newcommand{\ysovl}[3]{ #1, {Sov.\ Astron.\ Lett.,} {#2}, #3}
\newcommand{\yjetp}[3]{ #1, {Sov.\ Phys.\ JETP,} {#2}, #3}

\newcommand{\ymn}[3]{ #1, {MNRAS,} {#2}, #3}
\newcommand{\ynat}[3]{ #1, {Nature,} {#2}, #3}

\newcommand{\ysph}[3]{ #1, {Solar Phys.,} {#2}, #3}

\newcommand{\ypre}[3]{ #1, {Phys.\ Rev.\ E,} {#2}, #3}

\newcommand{\yjour}[4]{ #1, {#2}, {#3}, #4}

\newcommand{\ybook}[3]{ #1, {#2} (#3)}

\titlerunning{Magnetic flux concentrations in a polytropic atmosphere}
\authorrunning{I. R. Losada \etal}

\title{Magnetic flux concentrations in a polytropic atmosphere}

\author{I. R. Losada\inst{1,2}\and A. Brandenburg\inst{1,2}\and
N. Kleeorin\inst{3,1,4}\and I. Rogachevskii\inst{3,1,4}
}
\institute{
Nordita, KTH Royal Institute of Technology and Stockholm University,
Roslagstullsbacken 23, 10691 Stockholm, Sweden
\and
Department of Astronomy, AlbaNova University Center,
Stockholm University, 10691 Stockholm, Sweden
\and
Department of Mechanical Engineering, Ben-Gurion University of the Negev,
POB 653, Beer-Sheva 84105, Israel
\and
Department of Radio Physics, N.~I.~Lobachevsky State University of
Nizhny Novgorod, Russia
}
\date{\today,~ $ $Revision: 1.267 $ $}
\begin{document}

\abstract{
Strongly stratified hydromagnetic turbulence has recently been identified
as a candidate for explaining the spontaneous formation of magnetic flux
concentrations by the negative effective magnetic pressure instability
(NEMPI).
Much of this work has been done for isothermal layers, in which the
density scale height is constant throughout.
}{
We now want to know whether earlier conclusions regarding the size of
magnetic structures and their growth rates carry over to the case of polytropic
layers, in which the scale height decreases sharply as one approaches
the surface.
}{
To allow for a continuous transition from isothermal to polytropic
layers, we employ a generalization of the exponential function known as
the $q$-exponential.
This implies that the top of the polytropic layer shifts with changing
polytropic index such that the scale height is
always the same at some reference height.
We used both mean-field simulations (MFS) and direct numerical simulations (DNS)
of forced stratified turbulence to determine the resulting flux concentrations
in polytropic layers.
Cases of both horizontal and vertical applied magnetic fields were considered.
}{
Magnetic structures begin to form at a depth where the magnetic field
strength is a small fraction of the local equipartition field strength
with respect to the turbulent kinetic energy.
Unlike the isothermal case where stronger fields can give rise
to magnetic flux concentrations at larger depths, in the polytropic case the
growth rate of NEMPI decreases for structures deeper down.
Moreover, the structures that form higher up have a smaller
horizontal scale of about four times their local depth.
For vertical fields, magnetic structures of super-equipartition strengths
are formed, because such fields survive downward advection
that causes NEMPI with horizontal magnetic fields to reach premature
nonlinear saturation by what is called the ``potato-sack'' effect.
The horizontal cross-section of such structures found in DNS
is approximately circular, which is reproduced with MFS
of NEMPI using a vertical magnetic field.
}{
Results based on isothermal models can be applied locally
to polytropic layers.
For vertical fields, magnetic flux concentrations of super-equipartition
strengths form, which supports suggestions that sunspot formation
might be a shallow phenomenon.
}

\keywords{magnetohydrodynamics (MHD) -- hydrodynamics -- turbulence --
Sun: dynamo}

\maketitle

\section{Introduction}
\label{Introduction}

In a turbulent medium, the turbulent pressure can lead to dynamically
important effects.
In particular, a stratified layer can attain a density distribution
that is significantly altered compared to the nonturbulent case.
In addition, magnetic fields can change the situation further,
because it can locally suppress the turbulence and thus reduce the
total turbulent pressure (the sum of hydrodynamic and magnetic
turbulent contributions).
%On a length scale of enough turbulent eddies, this
%AB: their wording sounds weird now. How about:
\blue{On length scales encompassing many turbulent eddies}, this
total turbulent pressure reduction
must be compensated for by additional gas pressure, which can
lead to a density enhancement and thus to
horizontal magnetic structures that become
heavier than the surroundings and sink \citep{BKKMR11}.
This is quite the contrary of magnetic buoyancy, which is still expected
to operate on the smaller scale of magnetic flux tubes
and in the absence of turbulence.
Both effects can lead to instability: the latter is the magnetic
buoyancy or interchange instability
\citep{New61,Par66}, and the former is now often
referred to as negative effective magnetic pressure instability (NEMPI),
which has been studied at the level of mean-field theory for
the past two decades \citep{KRR89,KRR90,KMR93,KMR96,KR94,RK07}.
These are instabilities of a stratified continuous magnetic field,
while the usual magnetic buoyancy instability requires nonuniform and
initially separated horizontal magnetic flux tubes
\citep{Par55,Sp81,SCFM94}.

Unlike the magnetic buoyancy instability,
NEMPI occurs at the expense of turbulent energy
rather than the energy of the gravitational field.
The latter is the energy source of the magnetic
buoyancy or interchange instability.
NEMPI is caused by a negative turbulent
contribution to the effective mean magnetic
pressure (the sum of nonturbulent and turbulent
contributions). For large Reynolds numbers, this
turbulent contribution to the effective magnetic
pressure is larger than the nonturbulent one.
This results in the excitation of NEMPI and the
formation of large-scale magnetic structures --
even from an originally uniform mean magnetic field.

Direct numerical simulations (DNS) have recently
demonstrated the operation of NEMPI in
isothermally stratified layers
\citep{BKKMR11,KBKMR12b}. This is a particularly
simple case in that the density scale height is
constant; i.e., the computational burden of
covering large density variation is distributed
over the depth of the entire layer.
In spite of this simplification, it has been argued
that  NEMPI is important for explaining prominent features in the manifestation
of solar surface activity.
In particular, it has been associated with the formation of active
regions \citep{KBKMR12c,WLBKR13} and sunspots \citep{BKR13,Jab2}.
However, it is now important to examine the validity of conclusions
based on such simplifications using more realistic models.
In this paper, we therefore now consider a polytropic stratification,
for which the density scale height is smallest in the upper layers,
and the density variation therefore greatest.

NEMPI is a large-scale instability that can be excited in stratified
small-scale turbulence.
This requires (i) sufficient scale separation in the sense that
the maximum scale of turbulent motions, $\ell$, must be much smaller than
the scale of the system, $L$,
and (ii) strong density stratification such that the density scale
height $H_\rho$ is much smaller than $L$; i.e.,
\EQ
\ell\ll H_\rho\ll L.
\EN
However, both the size of turbulent motions and the typical size of
perturbations due to NEMPI can be related to the density scale height.
Furthermore, earlier work of \cite{KBKMR12c} using isothermal layers
shows that the scale of perturbations due to NEMPI
exceeds the typical density scale height.
Unlike the isothermal case, in which the scale height is constant, it decreases rapidly with height
in a polytropic layer.
It is then unclear how such structures could fit into the
narrow space left by the stratification and whether
the scalings derived for the isothermal case can still
be applied locally to polytropic layers.

NEMPI has already been studied previously for polytropic layers
in mean-field simulations (MFS) \citep{BKR10,KBKMR12,JBKR13},
but no systematic comparison has been made with NEMPI in isothermal
or in polytropic layers with different values of the polytropic index.
This will be done in the present paper, both in MFS and DNS.
Those two complementary types of simulations have proved to be a good
tool for understanding the underlying physics of NEMPI.
An example are the studies of the effects of rotation on NEMPI
\citep{Losada,Losada2}, where MFS have been able to give
quantitatively useful predictions before corresponding DNS
were able to confirm the resulting dependence.

\section{Polytropic stratification}

We discuss here the equation for the vertical profile of the fluid density
in a polytropic layer.
In a Cartesian plane-parallel layer with polytropic stratification,
the temperature gradient
is constant, so the temperature goes linearly to zero at $z_\infty$.
The temperature, $T$, is proportional to the square of the sound speed,
$\cs^2$, and thus also to the density scale height
$H_\rho(z)$, which is given by $H_\rho=\cs^2/g$
for an isentropic stratification, where
$\grav$ is the acceleration due to the gravity.
For a perfect gas,
the density $\rho$ is proportional to $T^n$, and the pressure $p$ is
proportional to $T^{n+1}$, such that $p/\rho$ is proportional to $T$,
where $n$ is the polytropic index.
Furthermore, we have $p(z)\propto\rho(z)^\Gamma$, where $\Gamma=(n+1)/n$
is another useful coefficient.

For a perfect gas, the specific entropy can be defined (up to an additive
constant) as $s=c_v\ln(p/\rho^\gamma)$,
where $\gamma=c_p/c_v$ is the ratio of specific heats
at constant pressure and constant density, respectively.
For a polytropic stratification, we have
\begin{equation}
\exp(s/c_v)=p/\rho^\gamma\propto\rho^{\Gamma-\gamma},
\end{equation}
so $s$ is constant when $\Gamma=\gamma$, which is
the case for an isentropic stratification.
In the following,  we make this assumption
and specify from now only the value of $\gamma$.
For a monatomic gas, we have $\gamma=5/3$, which
is relevant for the Sun, while for a diatomic
molecular gas, we have $\gamma=7/5$, which is relevant
for air. In those cases, a stratification with
$\Gamma=\gamma$ can be motivated by assuming
perfect mixing across the layer. The isothermal
case with $\gamma=1$ can be motivated by assuming
rapid heating/cooling to a constant temperature.

\begin{figure}[t!]\begin{center}
\includegraphics[width=\columnwidth]{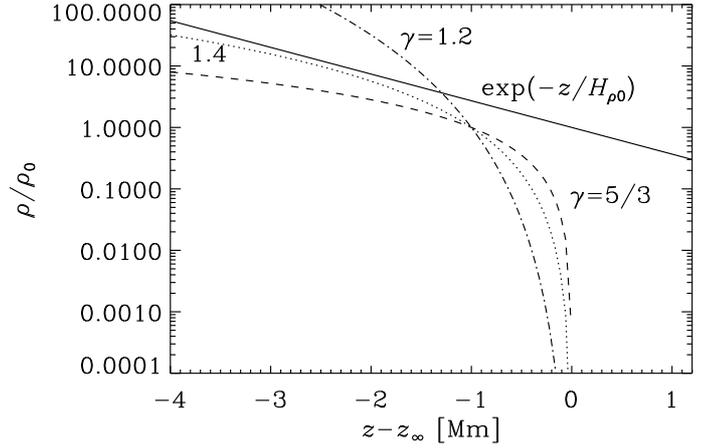}
\end{center}\caption[]{
Isothermal and polytropic relations for different values of $\gamma$
when calculated using the conventional formula
$\rho\propto(z_\infty-z)^n$ with $n=1/(\gamma-1)$.
}\label{ppoly_usual}\end{figure}

\begin{figure}[t!]\begin{center}
\includegraphics[width=\columnwidth]{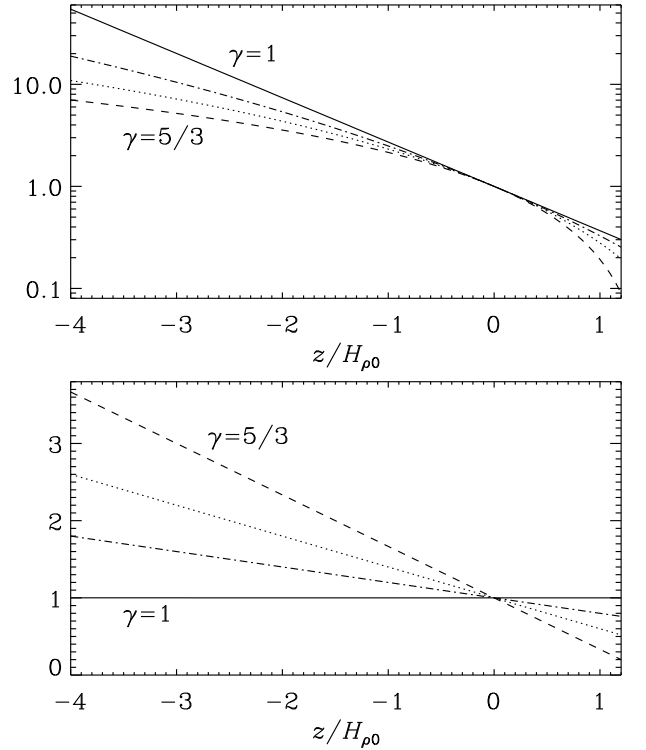}
\end{center}\caption[]{
Polytropes [Eq.~(\ref{rho_gen})] with $\gamma=1$ (solid line),
1.2 (dash-dotted), 1.4 (dotted), and 5/3 (dashed) and density scale height
[Eq.~(\ref{Hrho})] for $-4\leq -z/H_{\rho0}\leq1.2$.
The total density contrast is similar for $\gamma=1$ and 5/3.
}\label{ppoly}\end{figure}

Our aim is to study the
change in the properties of NEMPI in a continuous fashion as we go from an
isothermally stratified layer to a polytropic
one. In the latter case, the fluid density varies
in a power law fashion, $\rho\propto(z_\infty-z)^n$,
while in the former, it varies exponentially,
$\rho\propto\exp(z_\infty-z)$. This is shown in
\Fig{ppoly_usual} where we compare the
exponential isothermal atmosphere
with a family of polytropic atmospheres
with $\gamma=1.2$, 1.4, and 5/3.
Clearly, there is no continuous connection between
the isothermal case and the polytropic one in the limit
$\gamma\to1$.
This cannot be fixed by rescaling the isothermal density
stratification, because in \Fig{ppoly_usual} its values
would still lie closer to 5/3 than to 1.4 or 1.2.
Another problem with this description is that for polytropic solutions
the density is always zero at
$z=z_\infty$, but finite in the isothermal case.
These different behaviors between isothermal and polytropic
atmospheres can be unified by using the generalized exponential
function known as the ``$q$-exponential'' \citep[see, e.g.,][]{Yam02},
which is defined as
\begin{equation}
e_q(x)=\left[1+(1-q)x\right]^{1/(1-q)},
\label{EQ}
\end{equation}
where the parameter $q$ is related to $\gamma$
via $q=2-\gamma$.
This generalization of the usual exponential function was
originally introduced by \cite{Tsa88} in connection with a
possible generalization of the Boltzmann-Gibbs statistics.
Its usefulness in connection with stellar polytropes
has been employed by \cite{PP93}.
Thus, the density stratification is given by
\begin{equation}
\frac{\rho}{\rho_0}=\left[1+(\gamma-1)\left(
-{z\over H_{\rho0}}\right)\right ]^{1/(\gamma-1)}
=\left(1-{z\over nH_{\rho0}}\right)^n,\quad
\label{rho_gen}
\end{equation}
which reduces to $\rho/\rho_0=\exp(-z/H_{\rho0})$
for isothermal stratification with $\gamma\to1$ and $n\to\infty$.
The density scale height is then given by
\begin{equation}
H_{\rho}(z)=H_{\rho0}-(\gamma-1)z=H_{\rho0}-z/n.
\label{Hrho}
\end{equation}
In the following, we measure lengths in units of $H_{\rho 0}=H_\rho(0)$.

In \Fig{ppoly} we show the dependencies of $\rho(z)$ and $H_\rho$
given by \Eqs{rho_gen}{Hrho}
for different values of $\gamma$.
Compared with \Fig{ppoly_usual}, where $z_\infty$ is held fixed,
in \Fig{ppoly} it is equal to $z_\infty=nH_{\rho0}=H_{\rho0}/(\gamma-1)$.
The total density contrast is roughly the same in all
four cases for different $\gamma$, but for increasing values of $\gamma$, the vertical
density gradient becomes progressively stronger in the upper layers.

\begin{figure}[t!]\begin{center}
\includegraphics[width=\columnwidth]{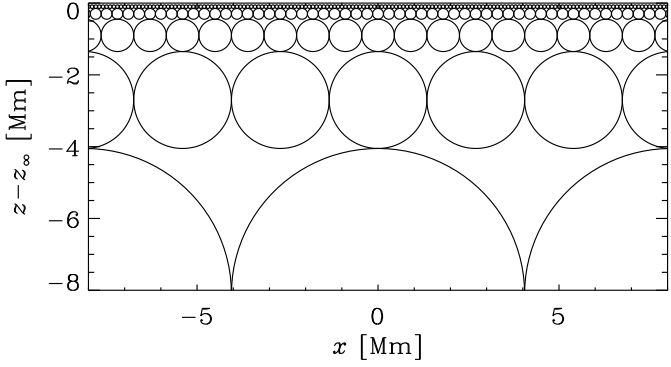}
\end{center}\caption[]{
Sketch showing the expected size distribution of the nearly circular
NEMPI eigenfunction structures at different heights.
}\label{psketch}\end{figure}

Assuming that the radius $R$ of the resulting structures
is proportional to $H_\rho$, we sketch in \Fig{psketch} a situation in
which $R$ is half the depth.
With $\rho\propto(z_\infty-z)^n$,
the density scale height is $H_\rho(z)=(z_\infty-z)/n$.
Thus, \Fig{psketch} applies to
a case in which $R=(z_\infty-z)/2=(n/2)\,H_\rho(z)$.
The solar convection zone is nearly isentropic and well described by $n=3/2$.
This means that the structures of \Fig{psketch} have $R=(3/4)\,H_\rho(z)$.
The results of \cite{KBKMR12c} and \cite{Jab2} suggest that
the horizontal wavenumber of structures, $k_\perp$,
formed by NEMPI, is less than or about
$H_\rho^{-1}$, so their horizontal wavelength is $\la2\pi H_\rho$.
One wavelength corresponds to the distance between two nodes,
i.e., the distance between two spheres, which is $4\,R$.
Thus, in the isothermal case we have $R/H_\rho=2\pi/4\approx1.5$, which implies
that such a structure would not fit into the isentropic atmosphere
described above.
This provides an additional motivation for our present work.

\section{DNS study}

In this section we study NEMPI in DNS for the polytropic layer.
Corresponding MFS are presented in \Sec{MFS}.

\subsection{The model}

We solve the equations for the magnetic vector potential $\AAA$,
the velocity $\UU$, and the density $\rho$, in the form
\begin{eqnarray}
{\partial\AAA\over\partial t}\!\!&=&\!\!\UU\times\BB-\eta\mu_0\JJ,\\
{\DD\UU\over\DD t}
\!\!&=\!\!&
{1\over\rho}\JJ\times\BB+\ff-\nu\QQ-\nab H,\\
{\DD\rho\over\DD t}\!\!&=&\!\!-\rho\nab\cdot\UU,
\end{eqnarray}
where $\DD/\DD t=\partial/\partial t+\UU\cdot\nab$ is the advective
derivative with respect to the actual (turbulent) flow,
$\BB=\BB_0+\nab\times\AAA$ is the magnetic field,
$\BB_0$  the imposed uniform field,
$\JJ=\nab\times\BB/\mu_0$  the current density,
$\mu_0$  the vacuum permeability,
$H=h+\Phi$  the reduced enthalpy, $h=c_pT$  the enthalpy,
$\Phi$ is the gravitational potential,
\EQ
-\QQ=\nabla^2\UU+\onethird\nab\nab\cdot\UU
+2\SSSS\nab\ln\rho
\EN
is a term appearing in the viscous force $-\nu\QQ$, 
$\SSSS$ is the traceless rate-of-strain tensor with components
\EQ
{\sf S}_{ij}=\half(\nabla_j U_i+\nabla_i U_j)-\onethird\delta_{ij}\nab\cdot\UU,
\EN
%\textcolor{red}{[Something is missing here, because you have 2 verbs. Also the 
%"and"\ before "S" means the list is coming to an end, but it doesn't. The same 
%occurs below between Eqs. 26 \&\ 27 where I suggest a correction that does not 
%begin any sentence with a symbol.] }
%IL: I don't understand which 2 verbs is he/she referring to 
%IL: this is repeated
%is the traceless rate-of-strain tensor of the flow,
%AB: The "traceless rate-of-strain tensor" came twice, and yes, there was a spurious "and"
$\nu$ is the kinematic viscosity,
and $\eta$ is the magnetic diffusion coefficient caused by electrical
conductivity of the fluid.
As in \cite{Losada}, $z$ corresponds to radius,
$x$ to colatitude, and $y$ to azimuth.
The forcing function $\ff$ consists of random, white-in-time,
plane, nonpolarized waves with a certain average wavenumber $\kf$.

\subsection{Boundary conditions and parameters}

In the DNS we use stress-free boundary conditions
for the velocity at the top and bottom; i.e.,
$\nabla_z U_{x}=\nabla_z U_{y}=U_z=0$.
For the magnetic field we use either perfect conductor boundary
conditions, $A_x=A_y=\nabla_z A_{z}=0$, or vertical field conditions,
$\nabla_z A_{x}=\nabla_z A_{y}=A_z=0$, again at both the top and bottom.
All variables are assumed periodic in the $x$ and $y$ directions.

\begin{figure*}[t!]\begin{center}
\includegraphics[width=\textwidth]{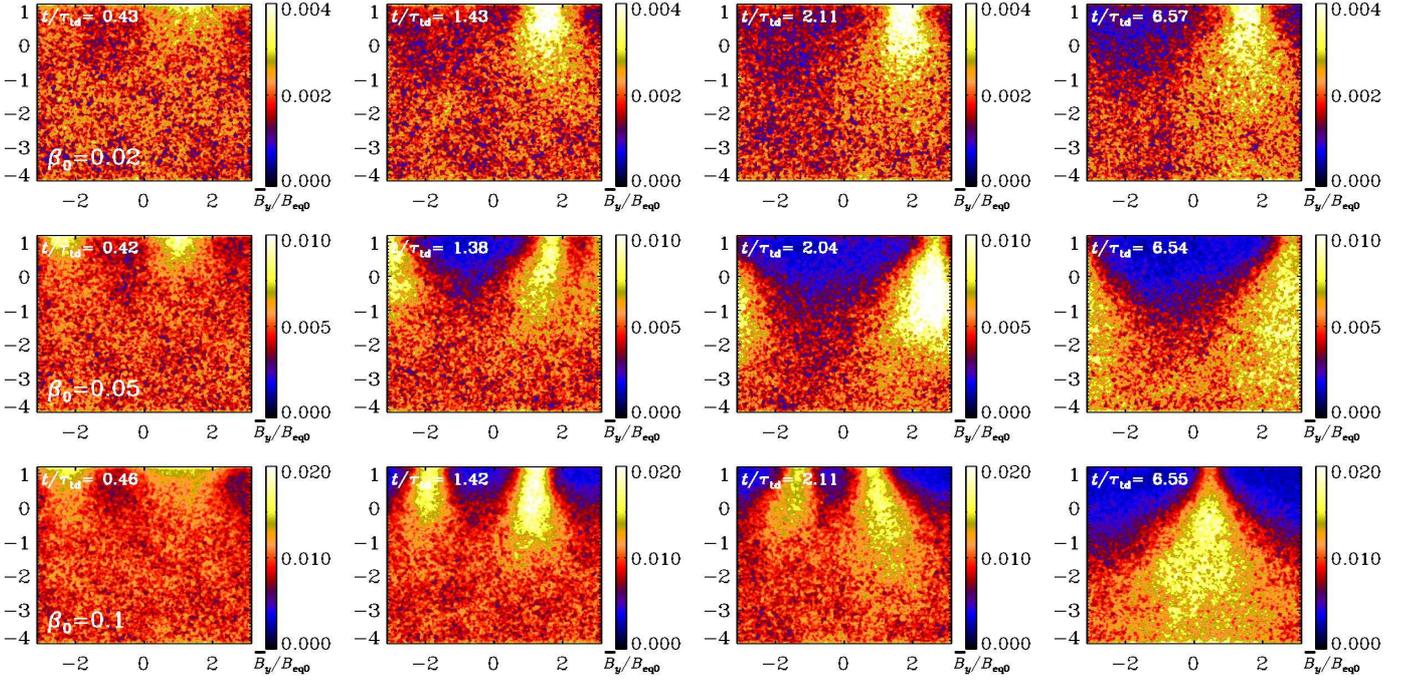}
\end{center}\caption[]{
Snapshots of $\meanB_y$ from DNS for $\gamma=5/3$ and
$\beta_0=0.02$ (upper row), $0.05$ (middle row), and $0.1$ (lower row)
at different times (indicated in turbulent-diffusive times, increasing
from left to right) in the presence of a horizontal field using the
perfect conductor boundary condition.
}\label{AB}\end{figure*}

The turbulent rms velocity is approximately
independent of $z$ with $\urms=\bra{\uu^2}^{1/2}\approx0.1\,\cs$.
The gravitational acceleration $\grav=(0,0,-g)$ is chosen such that
$k_1 H_{\rho0}=1$,
where $k_1=2 \pi/L$ and $L$ is the size of the domain.
With one exception (\Sec{MFScoefficients}), we always use the value $\kf/k_1=30 $
for the scale separation ratio.
For $\BB_0$ we choose either a horizontal field pointing in the $y$ direction
or a vertical one pointing in the $z$ direction.
The latter case, $\BB_0=(0,0,B_0)$, is usually combined with the use
of the vertical field boundary condition, while the former one,
$\BB_0=(0,B_0,0)$, is combined with using
perfect conductor boundary conditions.
The strength of the imposed field is expressed in terms of
$\Beqz=\Beq(z=0)$, which is the equipartition field strength at $z=0$.
Here, the equipartition field
$\Beq(z) = \left(\mu_0\meanrho(z)\right)^{1/2} \urms$.
The imposed field is normalized by $\Beqz$ and denoted as
$\beta_0=B_0/\Beqz$, while $\beta=|\meanBB|/\Beq$
is the modulus of the normalized mean magnetic field.
Time is expressed in terms of the turbulent-diffusive time,
$\tautd=H_{\rho0}^2/\etatz$, where $\etatz=\urms/3\kf$ \citep{SBS08}
is an estimate for the turbulent magnetic diffusivity used in the DNS.

Our values of $\nu$ and $\eta$ are characterized by specifying the
kinetic and magnetic Reynolds numbers,
\EQ
\Rey=\urms/\nu\kf,\quad
\Rm=\urms/\eta\kf.
\EN
In most of this paper (except in \Sec{MFScoefficients}) we use
$\Rey=36$ and $\Rm=18$, which are also the values used by \cite{KBKMR12c}.

The DNS are performed with the {\sc Pencil Code},
\url{http://pencil-code.googlecode.com}, which uses sixth-order explicit
finite differences in space and a third-order accurate time stepping method.
We use a numerical resolution of $256^3$ mesh points.

\begin{figure*}[t!]\begin{center}
\includegraphics[width=\textwidth]{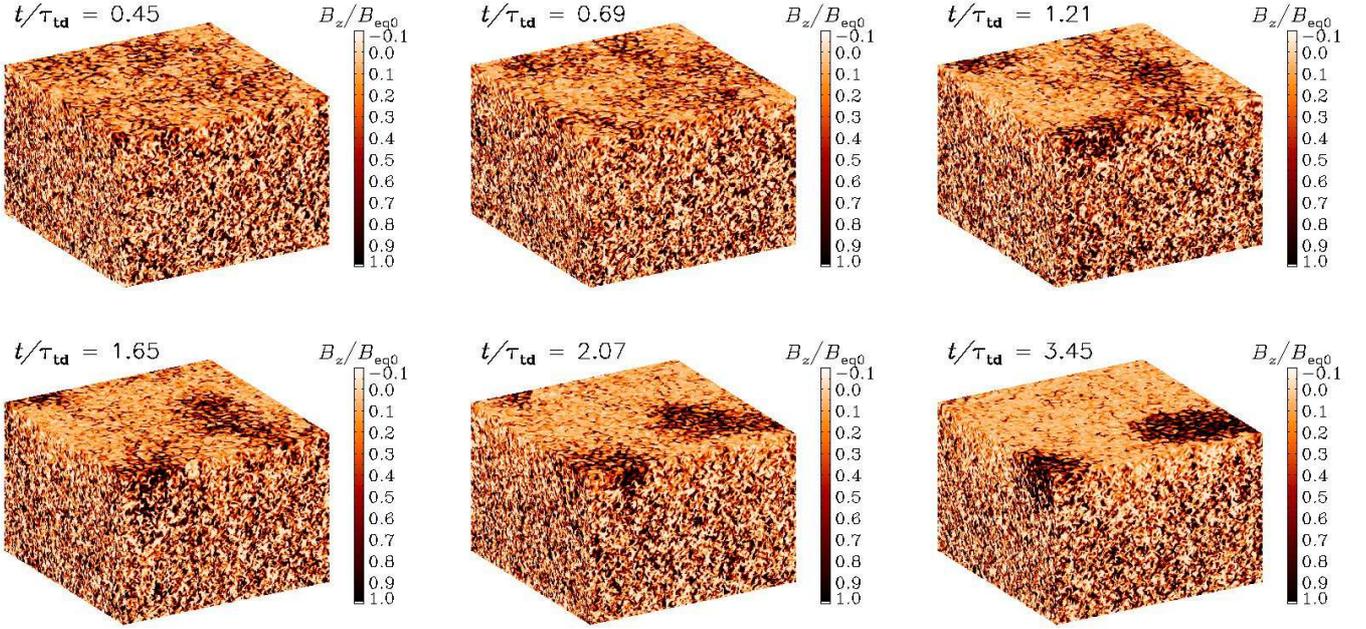}
\end{center}\caption[]{
Snapshots from DNS showing $\meanB_z$ on the periphery
of the computational domain for $\gamma=5/3$ and
$\beta_0=0.05$ at different times
for the case of a vertical field using the
vertical field boundary condition.
}\label{Vpoly}\end{figure*}

\begin{figure*}[t!]\begin{center}
\includegraphics[width=\textwidth]{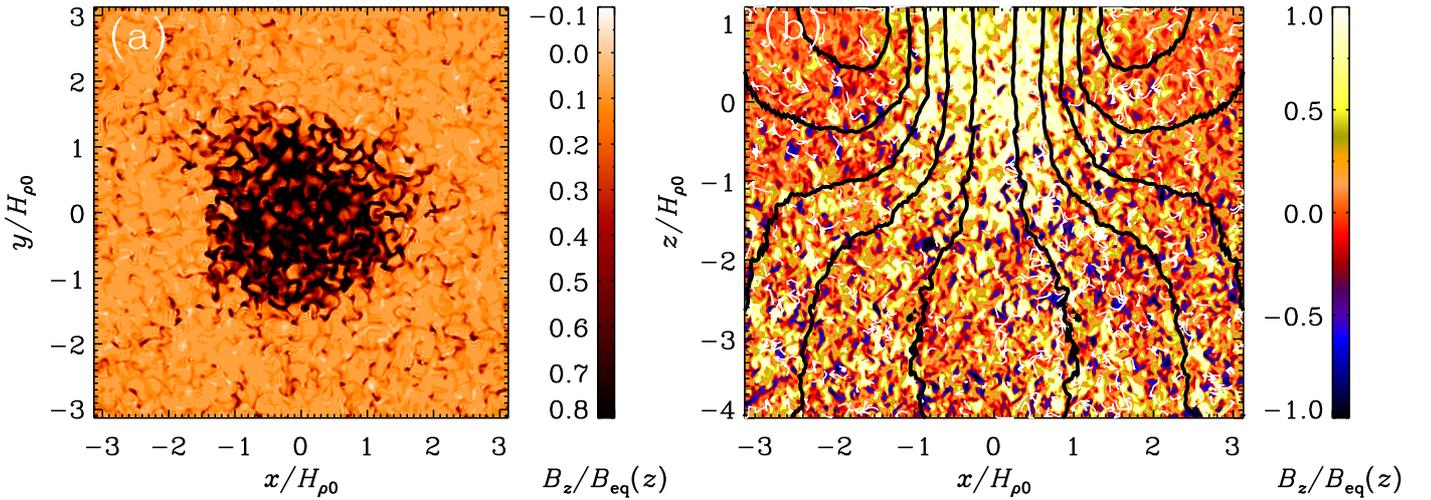}
\end{center}\caption[]{
Cuts of $B_z/\Beq(z)$ in (a) the $xy$ plane at the top boundary
($z/H_{\rho0}=1.2$) and (b) the $xz$ plane through the middle of the spot
at $y=0$ for $\gamma=5/3$ and $\beta_0=0.05$.
In the $xz$ cut, we also show magnetic field lines and flow vectors
obtained by numerically averaging in azimuth around the spot axis.
}\label{pslices_VV256gam53_B005b}\end{figure*}

\subsection{Horizontal fields}

We focus on the case $\gamma=5/3$ and show in \Fig{AB}
visualizations of $\meanB_y$ at different instants
for three values of the imposed
horizontal magnetic field strength.
For $\beta_0=0.02$, a magnetic structure is clearly visible at $t/\tautd=1.43$,
while for $\beta_0=0.05$ structures are already fully developed at
$t/\tautd=0.42$.
In that case ($\beta_0=0.05$), at early times ($t/\tautd=0.42$),
there are two structures, which then begin to merge at $t/\tautd=1.38$.
The growth rate of the magnetic structure is found to be
$\lambda\approx2\etatz/H_{\rho0}^2$ for the runs shown in \Fig{AB}.
This is less than the value of $\lambda\approx5\etatz/H_{\rho0}^2$
found earlier for the isothermal case \citep{KBKMR12b}.

For $\gamma=5/3$ and $\beta_0\leq0.02$, the magnetic structures become smaller
($k_\perp H_{\rho0}=2$) near the surface.
In the nonlinear regime, i.e., at late times, the structures
move downward due to the so-called ``potato-sack'' effect,
which was first seen in MFS
\citep{BKR10} and later confirmed in DNS \citep{BKKMR11}.
The magnetic structures sink in the nonlinear stage of NEMPI,
because an increase in the mean magnetic field inside the
horizontal magnetic flux tube increases the absolute
value of the effective magnetic pressure.
On the other hand, a decrease in
the negative effective magnetic pressure
is balanced out by increased gas pressure, which in turn leads
to higher density, so the magnetic structures become heavier
than the surroundings and sink.
This potato-sack effect has been clearly observed
in the present DNS with the polytropic layer
(see the right column in \Fig{AB}).

\subsection{Vertical fields}

Recent DNS using isothermal layers have shown that strong circular flux
concentrations can be produced in the case of a vertical
magnetic field \citep{BKR13,Jab2}.
This is also observed in the present study of a polytropic layer;
see \Fig{Vpoly}, where we show the evolution of $\meanB_z$ on the
periphery of the computational domain for $\gamma=5/3$ and $\beta_0=0.05$
at different times.
A difference to the DNS for $\gamma=1$ \citep{BKR13}
seems to be that for $\gamma=5/3$ the magnetic structures are
shallower than for $\gamma=1$; see \Fig{pslices_VV256gam53_B005b},
where we show $xy$ and $xz$ slices of $B_z$ through the spot.
Owing to periodicity in the $xy$ plane, we have shifted the
location of the spot to $x=y=0$.
We note also that the field lines of the averaged magnetic field
show a structure rather similar to the one found in MFS of \cite{Jab2}.
The origin of circular structures is associated with a cylindrical
symmetry for the vertical magnetic field.
The growth rate of the magnetic field in the spot is found to be
$\lambda\approx0.9\etatz/H_{\rho0}^2$, which is similar to the value of 1.3
found earlier for the isothermal case \citep{BKR13}.

\subsection{Effective magnetic pressure}
\label{MFScoefficients}

As pointed out in \Sec{Introduction}, the main reason for the formation
of strongly inhomogeneous large-scale magnetic structures
is the negative contribution
of turbulence to the large-scale magnetic pressure, so that
the effective magnetic pressure (the sum of turbulent and nonturbulent
contributions) can be negative at high magnetic Reynolds numbers.
The effective magnetic pressure has been determined from DNS
for isothermally stratified forced turbulence \citep{BKR10,BKKR12} and for
turbulent convection \citep{KBKMR12}.
To see whether the nature of polytropic stratification has any influence on
the effective magnetic pressure, we use DNS.

We first explain the essence of the effect of turbulence
on the effective magnetic pressure.
We consider the momentum equation in the form
\EQ
{\partial\over\partial t}\rho\, U_i=-{\partial\over\partial
x_j}\Pi_{ij} + \rho \, g_i,
\EN
where
\EQ
\Pi_{ij}=\rho \, U_iU_j+\delta_{ij}\left(p+\BB^2\!/2\mu_0\right)
-B_iB_j/\mu_0-2\nu \rho \, {\sf S}_{ij}
\label{Piorig}
\EN
is the momentum stress tensor and $\delta_{ij}$ 
the Kronecker tensor.

Ignoring correlations between velocity and density fluctuations
for low-Mach number turbulence, the averaged momentum
equation is
\EQ
{\partial\over\partial t} \meanrho \, \meanUU_i =
-{\partial\over\partial x_j}\overline{\Pi}_{ij} + \meanrho \, g_i,
\EN
where an overbar means $xy$ averaging,
$\overline\Pi_{ij}=\overline\Pi_{ij}^{\rm m}+\overline\Pi_{ij}^{\rm f}$
is the mean momentum stress tensor, split into contributions resulting
from the mean field (indicated by superscript m) and
the fluctuating field (indicated by superscript f).
The tensor $\overline\Pi_{ij}^{\rm m}$ has the same form as \Eq{Piorig},
but all quantities have now attained an overbar; i.e.,\
\EQ
\overline\Pi_{ij}^{\rm m}=\meanrho \, \meanU_i\meanU_j
+\delta_{ij}\left(\overline{p}+\meanBB^2\!/2\mu_0\right)
-\meanB_i\meanB_j/\mu_0-2\nu \meanrho \, \overline{\sf S}_{ij}.\;\;
\EN
The contributions, $\overline\Pi_{ij}^{\rm f}$, which
result from the fluctuations $\uu=\UU-\meanUU$ and
$\bb=\BB-\meanBB$ of velocity and magnetic fields, respectively,
are determined by
the sum of the Reynolds and Maxwell stress tensors:
\EQ
\overline\Pi_{ij}^{\rm f}=\meanrho \, \overline{u_iu_j}
+\delta_{ij}\overline{\bb^2}\!/2\mu_0-\overline{b_ib_j}/\mu_0.
\label{stress0}
\EN
This contribution, together with the contribution from the mean field,
$\overline\Pi_{ij}^{\rm m}$, comprises the total mean momentum tensor.
The contribution from the fluctuating fields is split into
a contribution that is independent of the mean magnetic field
$\overline\Pi_{ij}^{\rm f,0}$
(which determines the turbulent viscosity and background 
turbulent pressure)
and a contribution that depends on the mean magnetic field
$\overline\Pi_{ij}^{{\rm f},\overline{B}}$.
The difference between the two, $\Delta\overline\Pi_{ij}^{\rm f}=
\overline\Pi_{ij}^{{\rm f},\overline{B}}-\overline\Pi_{ij}^{\rm f,0}$,
is caused by the mean magnetic field and is parameterized in the form
\EQ
\Delta\overline\Pi_{ij}^{\rm f}= \mu_0^{-1} \left(q_{\rm s}\meanB_i\meanB_j
- q_{\rm p} \, \delta_{ij} \meanBB^2\!/2 - \qg \,
\ghat_i \, \ghat_j \meanBB^2\right),
\label{funct-tensor}
\EN
where the coefficient $\qp$ represents the isotropic
turbulence contribution to the mean magnetic pressure,
the coefficient $\qs$ represents the turbulence contribution to the
mean magnetic tension, while the coefficient $\qg$ is the anisotropic
turbulence contribution to the mean magnetic pressure,
and it characterizes the effect of vertical variations of the
magnetic field caused by the vertical magnetic pressure gradient.
Here, $\ghat_i$ is the unit vector in the direction of the gravity field
(in the vertical direction).
We consider cases with horizontal and vertical
mean magnetic fields separately.
Analytically, the coefficients $\qp$, $\qg,$ and $\qs$ have been
obtained using both the spectral $\tau$ approach \citep{RK07} and the
renormalization approach \citep{KR94}.
The form of \Eq{funct-tensor} is also obtained using simple symmetry arguments;
e.g., for a horizontal field, the linear
combination of three independent true tensors, $\delta_{ij}, \ghat_i \ghat_j$
and $\overline{B}_i \overline{B}_j$, yields \Eq{funct-tensor},
while for the vertical field, the linear
combination of two independent true tensors, $\delta_{ij}$
and $\overline{B}_i \overline{B}_j$, yields this ansatz.

Previous DNS studies \citep{BKKR12} have shown that $\qs$ and $\qg$ are
negligible for forced turbulence.
To avoid the formation of magnetic structures
in the nonlinear stage of NEMPI, which would
modify our results, we use here a lower scale separation ratio, $\kf/k_1=5$,
keeping $k_1 H_\rho=1$, and using $\Rey=140$ and $\Rm=70$,
as in \cite{BKKR12}.
To determine $\qp(\beta)$, it is sufficient to measure the
three diagonal components of $\overline\Pi_{ij}^{\rm f}$
both with and without an imposed magnetic field.

\begin{figure}[t!]\begin{center}
\includegraphics[width=\columnwidth]{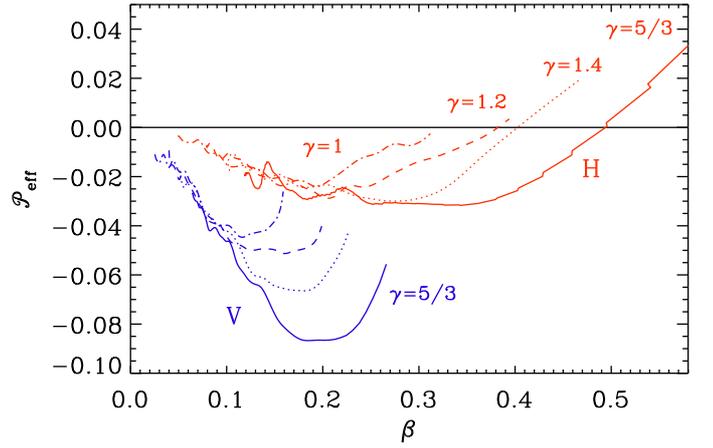}
\end{center}\caption[]{
Effective magnetic pressure obtained from DNS in a polytropic layer
with different $\gamma$ for horizontal (H, red curves)
and vertical (V, blue curves) mean magnetic fields.
}\label{pqps_new_comp}\end{figure}

\begin{figure}[t!]\begin{center}
\includegraphics[width=\columnwidth]{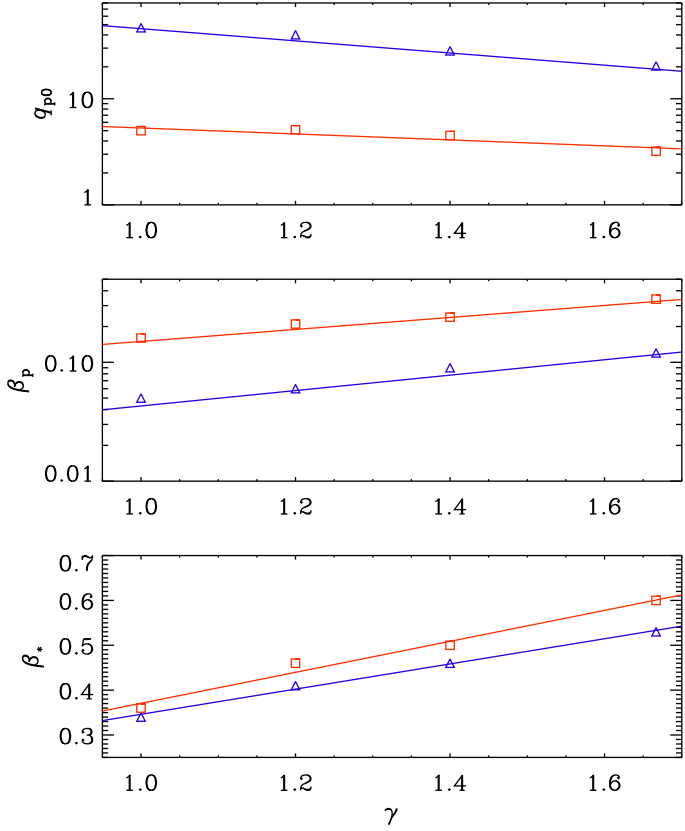}
\end{center}\caption[]{
Parameters $\qpz$, $\betap$, and
$\betastar$ for the function $\qp(\beta)$ [see \Eq{qp-apr}] versus
$\gamma$ for horizontal (red line)
and vertical (blue line) mean magnetic fields.
}\label{pfitcoeffs_res}\end{figure}

In \Fig{pqps_new_comp} we present the results for
forced turbulence in the polytropic layer
with different $\gamma$ for horizontal
and vertical mean magnetic fields.
It turns out that the normalized effective magnetic pressure,
\EQ
\Peff=\half(1-\qp)\beta^2,
\label{p-eff}
\EN
has a minimum value $\Pmin$ at $\betamin$.
Following \cite{KBKR12}, the function $\qp(\beta)$ is approximated by:
\EQ
\qp(\beta)={\qpz\over1+\beta^2/\betap^2}
={\betastar^2\over\betap^2+\beta^2},
\label{qp-apr}
\EN
where $\qpz$, $\betap$, and $\betastar=\betap\qpz^{1/2}$
are constants.
This equation can be understood as a quenching formula
for the effective magnetic pressure; see \cite{JBKR13}.
The coefficients $\betap$ and $\betastar$
are related to $\Pmin$ at $\betamin$ via \citep{KBKR12}
\begin{equation}
\betap=\betamin^2\left/\sqrt{-2\Pmin},\right.\;\;
\betastar=\betap+\sqrt{-2\Pmin}.
\end{equation}
In \Fig{pfitcoeffs_res} we show these fitting parameters for the
function $\qp(\beta)$ for polytropic layer with different $\gamma$
for horizontal and vertical mean magnetic fields.
The effects of negative effective magnetic pressure
are generally stronger for vertical magnetic fields
($\qpz$ is larger and $\betap$ smaller) than for
horizontal ones ($\qpz$ is smaller and $\betap$ larger),
but the values $\betastar=\betap\qpz^{1/2}$ are similar in both cases,
and increasing from 0.35 (for $\gamma=1$) to 0.6 (for $\gamma=5/3$);
see \Fig{pfitcoeffs_res}.

\section{Mean-field study}
\label{MFS}
We now consider two sets of parameters that we refer
to as Model I (with $\qpz=32$ and $\betap=0.058$ corresponding to
$\betastar=0.33$) and Model II ($\qpz=9$ and $\betap=0.21$
corresponding to $\betastar=0.63$).
These cases are representative of the strong (large $\betastar$)
and weak (small $\betastar$) effects of NEMPI.
Following earlier studies \citep{BKKR12},
we find $\qs$ to be compatible with zero.
We thus neglect this coefficient in the following.

\subsection{Governing parameters and estimates}
\label{estimates}

The purpose of this section is to summarize the
findings for the isothermal case in MFS.
One of the key results is the prediction of the growth rate
of NEMPI.
The work of \cite{KBKMR12c} showed that in the ideal case
(no turbulent diffusion), the growth rate $\lambda$ is
approximated well by
\begin{equation}
\lambda\approx\betastar\urms/H_{\rho}
\quad\mbox{(no turbulent diffusion)}.
\label{A1}
\end{equation}
However, turbulent magnetic diffusion, $\etat$, can clearly not
be neglected and is chiefly responsible for shutting off NEMPI
if the turbulent eddies are too big and $\etat$ too large.
This was demonstrated in Fig.~17 of \cite{BKKR12}.
A heuristic ansatz, which is motivated by similar circumstances
in mean-field dynamo theory \citep{KR80}, is to add a term
$-\etat k^2$ to the righthand side of Eq.~(\ref{A1}),
where $k$ is the wavenumber of NEMPI.

To specify the expression for $\lambda$, we normalize
the wavenumber of the perturbations
by the inverse density scale height
and denote this by $\kappa\equiv k H_\rho$.
The wavenumber of the energy-carrying turbulent eddies $\kf$ is in
nondimensional form $\kappaf\equiv\kf H_\rho$, and the normalized
horizontal wavenumber of the resulting magnetic structures is referred to
as $\kappa_\perp=k_\perp H_\rho$.
For NEMPI, these values have been estimated to be
$\kappa_\perp=0.8$--1.0, and can even be smaller
for vertical magnetic fields \citep{Jab2}.
Using an approximate aspect ratio of unity for magnetic structures,
we have $\kappa=\sqrt{2}\kappa_\perp\approx1.1$--1.4.
In stellar mixing length theory \citep{Vit53}, the mixing length is
$\ell_{\rm f}=\alpha_{\rm mix} H_\rho$, where $\alpha_{\rm mix}\approx1.6$
is a nondimensional mixing length parameter.
Since $\kf= 2 \pi/\ell_{\rm f} $, we arrive at the following estimate:
$\kappaf=2\pi\gamma/\alpha_{\rm mix}\approx6.5$.
[Owing to a confusion between pressure and density scale heights,
this value was underestimated by \cite{KBKMR12c} to be $2.4$,
although an independent calculation of this value from turbulent
convection simulations would still be useful.]
Using $\etat\approx\urms/3\kf$, the turbulent magnetic diffusive rate
for an isothermal atmosphere is given by
\begin{equation}
\etat k^2={\urms\over3H_\rho}{\kappa^2\over\kappaf},
\label{etatk2}
\end{equation}
and the growth rate of NEMPI in that normalization is
\begin{equation}
{\lambda\over\etat k^2}=3\betastar{\kappaf\over\kappa^2}-1.
\label{lambda}
\end{equation}
Using $\betastar=0.23$, which is the relevant value for high
magnetic Reynolds numbers \citep{BKKR12}, we find
$\lambda/\etat k^2\approx2.7$--$1.3$.
However, since \Fig{pfitcoeffs_res} shows an increase of
$\betastar$ with increasing polytropic index, one might expect
a corresponding increase in the growth rate of NEMPI
for a polytropic layer, in which $H_\rho$ varies strongly
with height $z$.
Indeed, in a polytropic atmosphere, $H_\rho$ is proportional to depth.
Thus, at any given depth there is a layer beneath,
where the stratification is less strong and
the growth rate of NEMPI is lower.
In addition, there is a thinner, more strongly stratified layer above,
where NEMPI might grow faster if only the
structures generated by NEMPI have enough room to
develop before they touch the top of the atmosphere at $z_\infty$,
where the temperature vanishes.

\subsection{Mean-field equations}

In the following, we consider MFS and compare it with DNS.
We also compare our MFS results with those of DNS \citep{BKKMR11,KBKMR12c}
using a similar polytropic setup.

The governing equations for the mean quantities (denoted by an overbar)
are fairly similar to those for the original equations, except that
in the MFS viscosity and magnetic diffusivity are replaced by their
turbulent counterparts, and the mean Lorentz force is supplemented
by a parameterization of the turbulent contribution
to the effective magnetic pressure.

The evolution equations for mean vector potential $\meanAA$,
mean velocity $\meanUU$, and mean density $\meanrho$, are
\EQA
\label{dAmean}
{\partial\meanAA\over\partial t}&=&\meanUU\times\meanBB
-\etaT\mu_0\meanJJ,\\
\label{dUmean}
{\meanDD\,\meanUU\over\meanDD t}
&=&{1\over\meanrho}\left[
\meanJJ\times\meanBB+\nab(q_{\rm p}\meanBB^2/2\mu_0)\right]
-\nuT\meanQQ-\nab\meanH,\\
{\meanDD\,\meanrho\over\meanDD t}&=&-\meanrho\nab\cdot\meanUU,
\ENA
where $\meanDD/\meanDD t=\partial/\partial t+\meanUU\cdot\nab$
is the advective derivative with respect to the mean flow,
$\meanrho$  the mean density,
$\meanH=\meanh+\Phi$  the mean reduced enthalpy,
$\meanh=c_p\meanT$  the mean enthalpy,
$\meanT\propto\meanrho^{\gamma-1}$  the mean temperature,
$\Phi$  the gravitational potential,
$\etaT=\etat+\eta$, and $\nuT=\nut+\nu$ are the sums of turbulent and
microphysical values of magnetic diffusivity and kinematic viscosities,
respectively. Also,
$\meanJJ=\nab\times\meanBB/\mu_0$  is the mean current density,
$\mu_0$ is the vacuum permeability,
\EQ
-\meanQQ=\nabla^2\meanUU+\onethird\nab\nab\cdot\meanUU
+2\meanSSSS\nab\ln\meanrho
\EN
is a term appearing in the mean viscous force $-\nuT\meanQQ$,
where $\meanSSSS$ is the traceless rate-of-strain tensor of the mean flow
with components $\meanSSS_{ij}=\half(\nabla_j\meanU_{i}+\nabla_i\meanU_{j})
-\onethird\delta_{ij}\nab\cdot\meanUU$,
and finally the term $\nab(q_{\rm p}\meanBB^2/2\mu_0)$
on the righthand side of Eq.~(\ref{dUmean})
determines the turbulent contribution to the
effective magnetic pressure.
Here, $q_{\rm p}$ depends on the local field strength;
see Eq.~(\ref{qp-apr}).
This term enters with a plus sign, so positive values of $\qp$
correspond to a suppression of the total turbulent pressure.
The net effect of the mean magnetic field leads to an effective
mean magnetic pressure that becomes negative for $\qp>1$.
This can indeed be the case for magnetic Reynolds numbers well
above unity \citep{BKKR12};
see also \Fig{pqps_new_comp} for a polytropic atmosphere.

The boundary conditions for MFS are the same as for DNS, i.e.,
stress-free for the mean velocity at the top and bottom.
For the mean magnetic field, we use either perfect conductor
boundary conditions (for horizontal, imposed magnetic fields)
or vertical field conditions (for vertical, imposed fields)
at the top and bottom.
All mean-field variables are assumed to be periodic in the $x$ and $y$ 
directions.
The MFS are performed again with the {\sc Pencil Code}, which is equipped
with a mean-field module for solving the corresponding equations.

\subsection{Expected vertical dependence of NEMPI}

\begin{figure}[t!]\begin{center}
\includegraphics[width=\columnwidth]{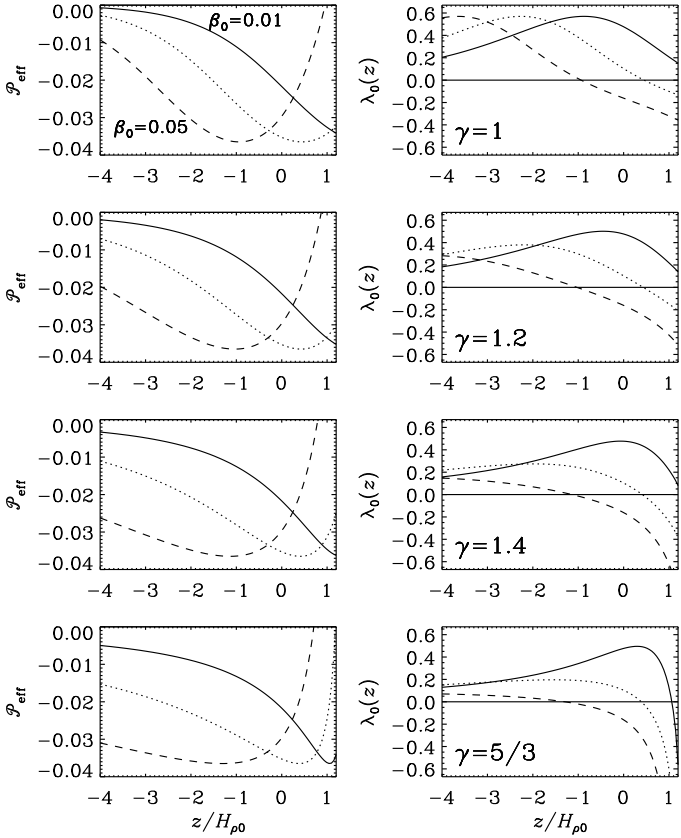}
\end{center}\caption[]{
Comparison of $\Peff(z)$ and $\lambda_0(z)$ in MFS for polytropic layers
with four values of $\gamma$ (top to bottom) and three values of $\beta$
(different line types).
}\label{ppeff}\end{figure}

To get an idea about the vertical dependence of NEMPI, we now consider
the resulting dependencies of $\Peff(z)$; see the lefthand
panels of \Fig{ppeff}.
We note that $\Peff$ is just a function of $\beta$ (\Eqs{p-eff}{qp-apr}),
which allows us to approximate the
local growth rates as \citep{RK07,KBKMR12c}
\begin{equation}
\lambda_0={\vA\over H_{\rho0}}
\left(-2{\dd\Peff\over\dd\beta^2}\right)^{\half},
\label{lam0}
\end{equation}
which are plotted in the righthand panels of \Fig{ppeff} for Model~I.

\subsection{Horizontal fields}

To analyze the kinematic stage of MFS, we measure
the value of the maximum downflow speed, $|\meanU|^{\rm down}_{\max}$ at each height.
We then determine the time interval during which the
maximum downflow speed increases exponentially and when the
height of the peak is constant and equal to $z_B$.
This yields the growth rate of the instability as
$\lambda=\dd\ln|\meanU|^{\rm down}_{\max}/\dd t$.

\begin{figure}[t!]\begin{center}
\includegraphics[width=\columnwidth]{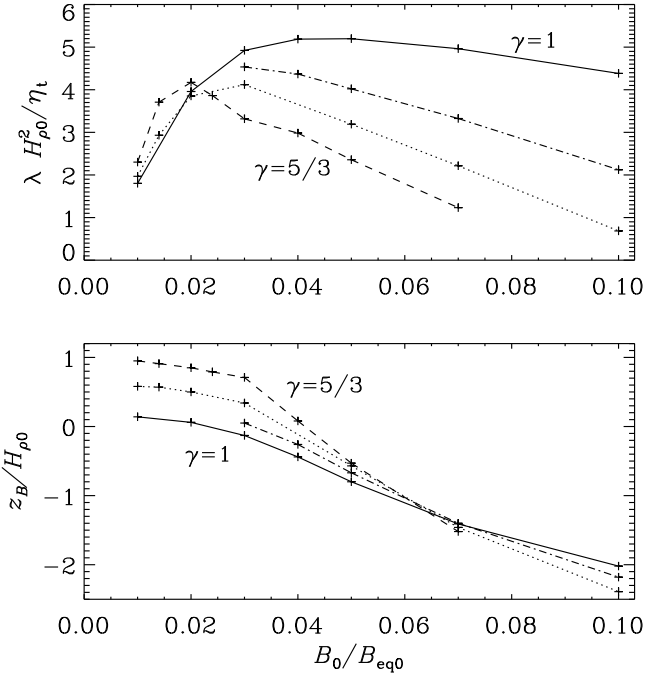}
\end{center}\caption[]{
Dependence of growth rate and
height where the eigenfunction attains its maximum value
(the optimal depth of NEMPI) on field strength from MFS
for different values of $\gamma$
in the presence of a horizontal field for Model~I.
}\label{pumax_gam}\end{figure}

In \Figs{pumax_gam}{pumax_gam_qp9} we plot, respectively for Models~I and II,
$\lambda$ (in units of $\etat/H_{\rho 0}^2$) and $z_B$
versus horizontal imposed magnetic field strength, $B_0/\Beq$.
The maximum growth rates for $\gamma=5/3$ and $\gamma=1$
are similar for both models (4--5\,$\etat/H_{\rho 0}^2$ for Model~I
and 16--20\,$\etat/H_{\rho 0}^2$ for Model~II).
It turns out that for $\gamma=5/3$, the growth rate $\lambda$
attains a maximum at some value
$B_0=B_{\max}$, and then it decreases with increasing $B_0$,
while in an isothermal run $\lambda$ is nearly constant
for greater field strength,
except near the surface where the proximity to the boundary is too small.
This close proximity reduces the growth rate.
For Model~I with $\gamma=1$, the decline of $\lambda$
(toward weaker fields on the lefthand side of the plot) begins when
the distance to the top boundary ($z_{\rm top}-z_B\approx1.2\,H_{\rho0}$
for $\beta_0=0.02$) is less than the radius of magnetic structures
($R\approx2\pi/k_\perp\approx1.5\,H_{\rho0}$ using $k_\perp H_{\rho0}=1$).
In Model~II with $\gamma=1$, the decline of $\lambda$ occurs for stronger
fields, but the distance to the top boundary ($\approx1.0\,H_{\rho0}$)
is still nearly the same as for Model~I.

\begin{figure}[t!]\begin{center}
\includegraphics[width=\columnwidth]{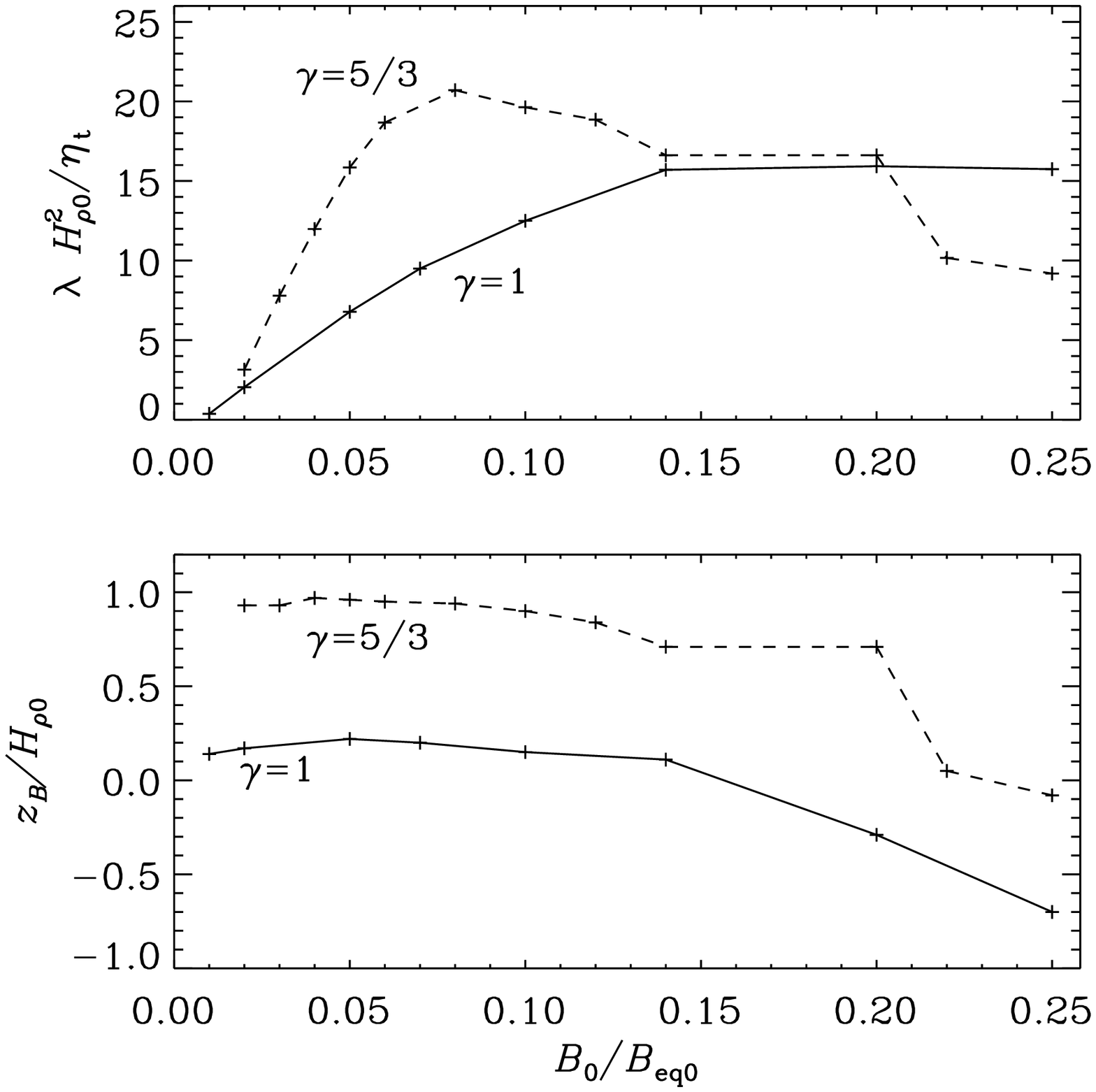}
\end{center}\caption[]{
Same as \Fig{pumax_gam} (horizontal field), but for Model~II.
}\label{pumax_gam_qp9}\end{figure}

\begin{figure}[t!]\begin{center}
\includegraphics[width=\columnwidth]{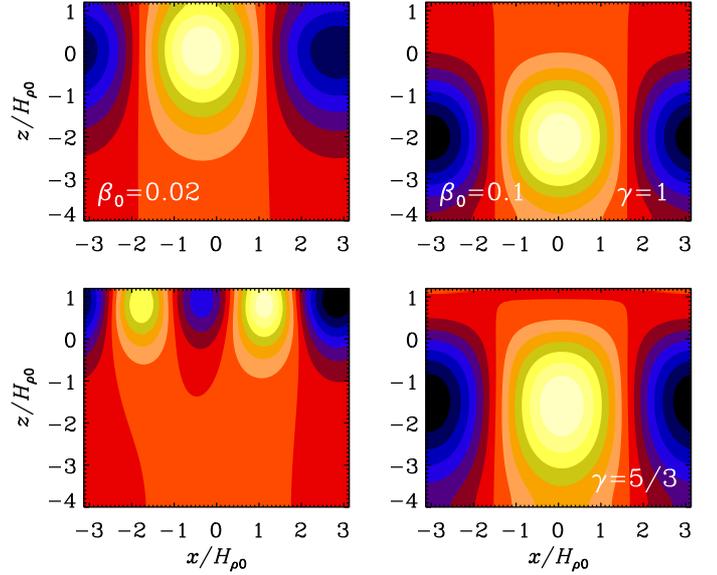}
\end{center}\caption[]{
Snapshots of $\meanB_y$ from MFS during the kinematic growth phase
for different values of $\gamma$ and $B_0/\Beqz$
for $\gamma=1$ (top) and $5/3$ (bottom) with $\beta_0=0.02$ (left)
and 0.1 (right) in the presence of a horizontal field for Model I.
}\label{psnap}\end{figure}

In an isothermal layer, the height where the eigenfunction peaks
is known to decrease with increasing field strength;
see Figure~6 of \cite{KBKR12}.
One might have expected this decrease to be less steep in the
polytropic case, because the optimal depth where NEMPI occurs
cannot easily be decreased without suffering a dramatic decrease
of the growth rate.
This is however not the case, and we find
that the optimal depth of NEMPI is
now falling off more quickly in Model~I, but is more similar
for Model~II; see the second panels of \Figs{pumax_gam}{pumax_gam_qp9}.
This means that in a polytropic layer, NEMPI works more effectively,
and its growth rate is fastest when the magnetic field is not too strong.
At the same time, the optimal depth of NEMPI increases, i.e.,
the resulting value of $z_B$ increases as $B_0$ decreases.

The resulting growth rates are somewhat less for Model~I
and somewhat higher for Model~II than those
of earlier mean-field calculations of \cite{KBKR12} (their Fig.~6),
who found $\lambda\Hpz^2/\etat\approx9.7$ in a model with
$\betastar=0.32$ and $\betap=0.05$.
These differences in the growth rates are plausibly explained by
differences in the mean-field parameters.

Visualizations of the resulting horizontal field structures are shown
in \Fig{psnap} for two values of $\gamma$ and $B_0$.
Increase in the parameter $\gamma$ results in a stronger localization
of the instability at the surface layer, where the density scale height
is minimum and the growth of NEMPI is strongest.

\subsection{Vertical fields}

In the presence of a vertical field, the early evolution of the instability
is similar to that for a horizontal field.
In both cases, the maximum field strength occurs at a somewhat larger
depth when saturation is reached, except that shortly before saturation
there is a brief interval during which the location of maximum field
strength rises slightly upwards in the vertical field case.
In the saturated case, however, the flux concentrations from NEMPI
are much stronger compared to the case of a horizontal field
and it leads to the formation of magnetic flux concentrations of
equipartition field strength \citep{BKR13,Jab2}.
This is possible because the resulting vertical flux tube
is not advected downward with
the flow that develops as a consequence of NEMPI.
The latter effect is the aforementioned ``potato-sack'' effect,
which acts as a nonlinear saturation mechanism of NEMPI
with a horizontal field.

\begin{figure}[t!]\begin{center}
\includegraphics[width=\columnwidth]{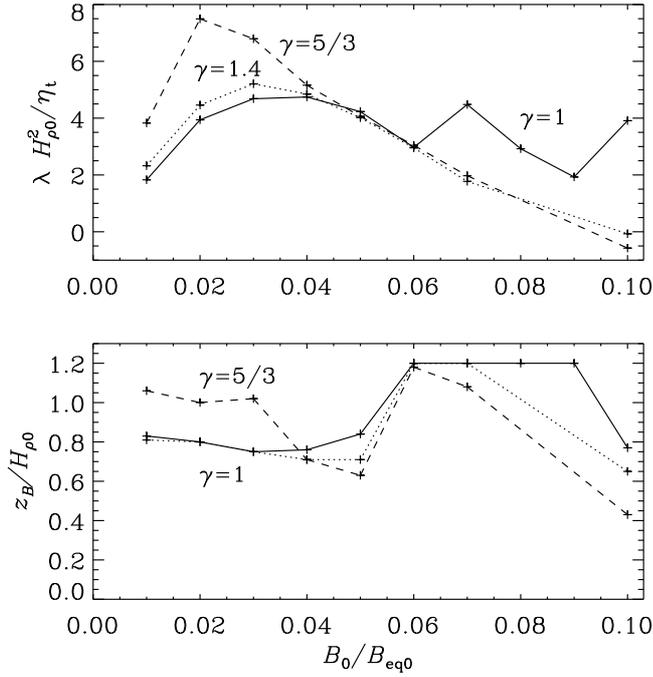}
\end{center}\caption[]{
Dependence of growth rate and optimal depth of NEMPI
on field strength from MFS for different values of $\gamma$
in the presence of a vertical field for Model~I.
}\label{pumax_Vgam}\end{figure}

In \Figs{pumax_Vgam}{pumax_Vgam_qp9} we plot the growth rates $\lambda$
and the heights where the eigenfunction attains its maximum values
for different $\beta_0=B_0/\Beqz$ for Models~I and II, respectively.
For $\gamma=5/3$, the maximum growth rate is higher larger than for $\gamma=1$.
This is true for Models~I and II, where they are
8--10\,$\etat/H_{\rho 0}^2$ for $\gamma=5/3$ and
5--7\,$\etat/H_{\rho 0}^2$ for $\gamma=1$.
The nonmonotonous behavior seen in the dependence of $\lambda$
on $B_0$ is suggestive of the presence of different mode structures,
although a direct inspection of the resulting magnetic field did not
show any obvious differences.
However, this irregular behavior may be related to artifacts resulting
from a finite domain size and were not regarded important enough
to justify further investigation.

\begin{figure}[t!]\begin{center}
\includegraphics[width=\columnwidth]{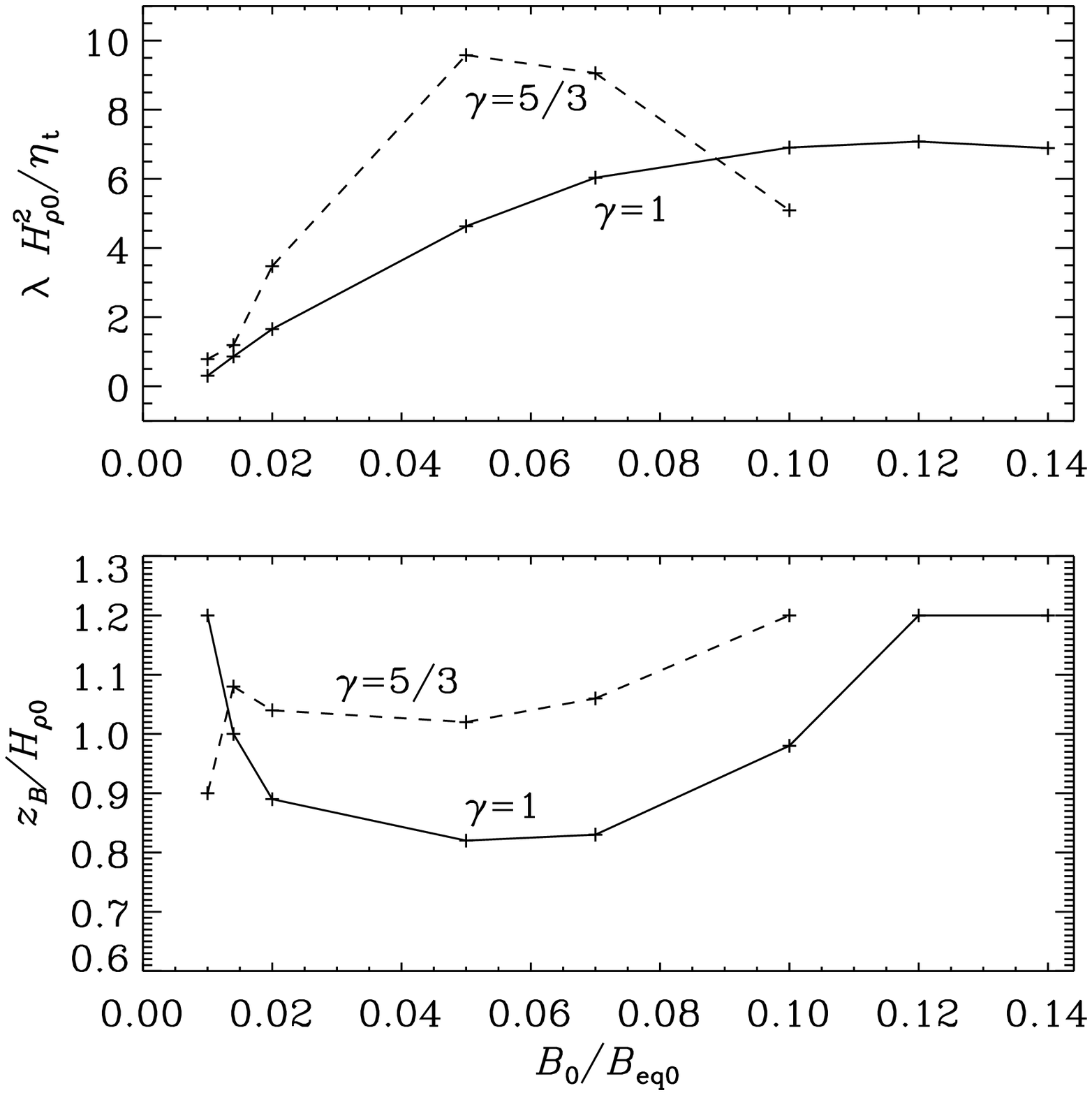}
\end{center}\caption[]{
Same as \Fig{pumax_Vgam} (vertical field), but for Model~II.
}\label{pumax_Vgam_qp9}\end{figure}

\begin{figure}[t!]\begin{center}
\includegraphics[width=\columnwidth]{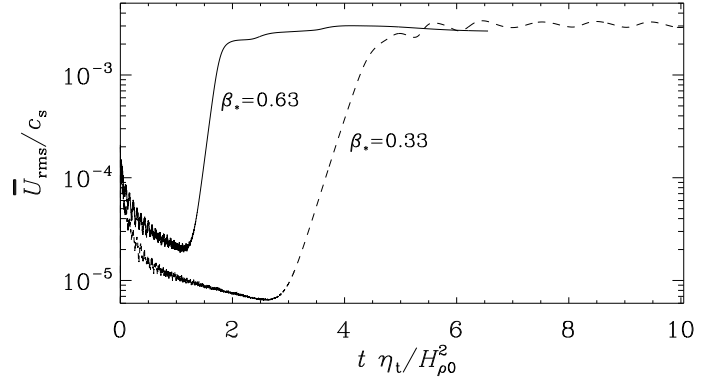}
\end{center}\caption[]{
Comparison of $\meanU_{\rm rms}$ from MFS for Models~I
(solid line) and II (dashed line), showing exponential
growth followed by nonlinear saturation.
In both cases we have $\gamma=5/3$ and an imposed vertical
magnetic field with $\beta_0=0.05$.
}\label{pcomp_VV3D64x64x128gam53_B005a}\end{figure}

Next we focus on a comparison of the growth rates
obtained from MFS for horizontal and vertical fields.
The values of $\beta_0$, $z_B$, and $\beta(z_B)=B_0/\Beq(z_B)$
for horizontal and vertical fields are compared in \Tab{Toptimal}
for both models.
We see that NEMPI is most effective in regions where the mean
magnetic field is a small fraction of the local equipartition field
and typically slightly less for $\gamma=5/3$ than for $\gamma=1$.
Indeed, for Model~I, $\beta(z_B)$ is 3--4\% for horizontal fields
and 3--6\% for vertical fields, while for Model~II,
$\beta(z_B)$ is 12--17\% for horizontal fields and
8--22\% for vertical fields.
Here, we have used $\beta(z_B)=\beta_0 e_{2-\gamma}(z_B/2H_{\rho0})$,
where $e_q(x)$ is the $q$-exponential function
defined by \Eq{EQ}.

\begin{table}[b!]\caption{
Comparison of the optimal depth $z_B$ and the corresponding normalized magnetic
field strength $\beta(z_B)$ for three values of $\gamma$
for imposed horizontal and vertical magnetic fields
of normalized strengths $\beta_0=B_0/\Beqz$, for Model~I.
}
\vspace{12pt}\centerline{\begin{tabular}{cc|crl|crl}
& & \multicolumn{3}{c|}{horizontal field} &
\multicolumn{3}{c}{vertical field} \\
Mod& $\gamma$ & $\beta_0$ & $\!\!z_B/H_{\rho0}\!\!$ & $\beta(z_B)$
         & $\beta_0$ & $\!\!z_B/H_{\rho0}\!\!$ & $\beta(z_B)$ \\
\hline
I & 1   & 0.05 & $-0.76$ & 0.034 & 0.04 & $0.76$ & 0.058 \\
I & 1.4 & 0.03 & $ 0.34$ & 0.035 & 0.03 & $0.75$ & 0.043 \\
I & 5/3 & 0.02 & $ 0.85$ & 0.029 & 0.02 & 1.0~~~  & 0.031 \\
\hline
II & 1   & 0.20 & $-0.29$ & 0.17 & 0.12 & 1.2~~~ & 0.22 \\
II & 5/3 & 0.08 & $ 0.94$ & 0.12 & 0.05 & 1.0~~~ & 0.08
\label{Toptimal}\end{tabular}}\end{table}

We expect that higher values of $\betastar$ will lead to greater growth rates.
To verify this, we compare in \Fig{pcomp_VV3D64x64x128gam53_B005a} the
time evolutions of $\meanU_{\rm rms}$ for Models~I (with $\betastar=0.33$)
and Model~II (with $\betastar=0.63$).
The growth rate has now increased by a factor of 2.4
(from $\lambda H_{\rho0}^2/\etat=3.9$ to 9.5),
which is slightly more than what is expected from $\betastar$,
which has increased by a factor of 1.9.
This change in the growth rate can also be seen in \Fig{pumax_gam_qp9} and
\Fig{pumax_Vgam_qp9} for Model~II (in comparison with
\Figs{pumax_gam}{pumax_Vgam} for Model~I).
The dependence of the growth rate on the magnetic field strength
is qualitatively similar for Models~I and II.
In particular, it becomes constant for $\gamma=1$,
but declines for $\gamma=5/3$ as the field increases.
The increase of $z_B$ with $B_0$ is, however, less strong for Model~II.

\begin{figure*}[t!]\begin{center}
\includegraphics[width=\textwidth]{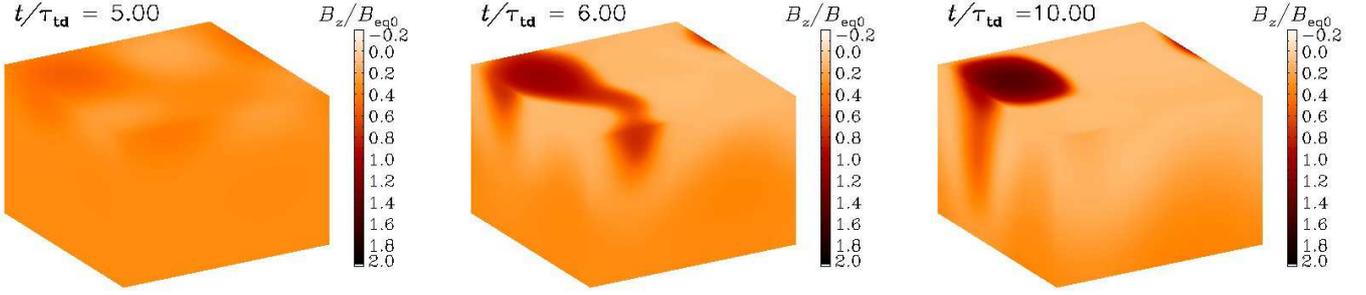}
\end{center}\caption[]{
Snapshots from MFS showing $\meanB_z$ on the periphery
of the computational domain for $\gamma=5/3$ and
$\beta_0=0.05$ at different times for Model I
for the case of a vertical field.
}\label{VpolyMFS}\end{figure*}

\begin{figure*}[t!]\begin{center}
\includegraphics[width=\textwidth]{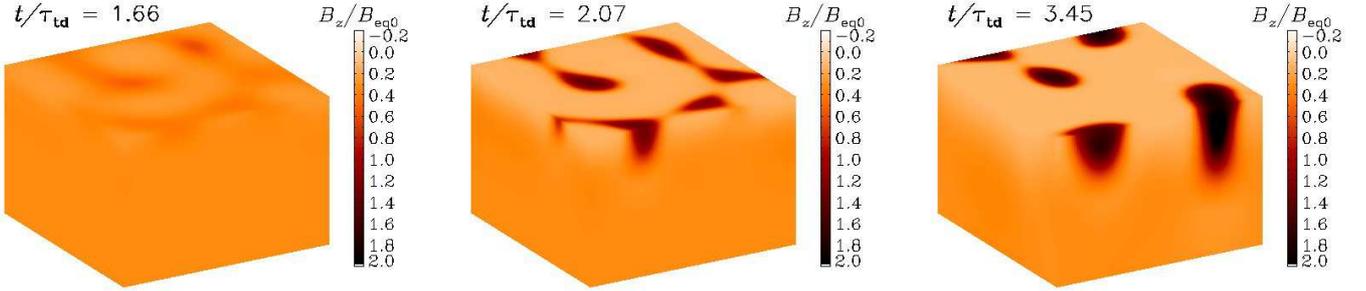}
\end{center}\caption[]{
Similar to \Fig{VpolyMFS} of MFS, but for Model II
at times similar to those in the DNS of \Fig{Vpoly}.
There are now more structures than in the earlier MFS
of \Fig{VpolyMFS}, and they develop more rapidly.
}\label{VpolyMFSii}\end{figure*}

Snapshots of $\meanB_z$ from MFS for $\gamma=5/3$ and
$\beta_0=0.05$ at different times
for Model~I are shown in \Fig{VpolyMFS}.
Comparison with the results of MFS for Model~II
(see \Fig{VpolyMFSii}) shows that Model~II fits the DNS better.
This is also seen by comparing \Fig{VpolyMFSii} with \Fig{Vpoly}.
However, our basic conclusions formulated in this paper are not affected.

\section{Conclusions}

The present work has demonstrated that in a polytropic layer,
both in MFS and DNS,
NEMPI develops primarily in the uppermost layers,
provided the mean magnetic field is not too strong.
If the field gets stronger, NEMPI can still develop,
but the magnetic structures now occur at greater depths
and the growth rate of NEMPI is lower.
However, at some point when the magnetic field gets too strong,
NEMPI is suppressed in the case of a polytropic layer,
while it would still operate in the isothermal case,
provided the domain is deep enough.
The slow down of NEMPI is not directly a consequence of
a longer turnover time at greater depths, but it is related
to stratification being too weak for NEMPI
to be excited.

By and large, the scaling relations determined previously
for isothermal layers with constant scale height still seem
to apply locally to polytropic layers with variable scale heights.
In particular, the horizontal scale of structures was previously
determined to be about $6$--$8\,H_\rho$ \citep{KBKMR12c,Jab2}.
Looking now at \Fig{AB}, we see that for $\beta_0=0.02$ and
$\gamma=5/3$, the structures have a wavelength of $\approx3\,H_{\rho0}$,
but this is at a depth where $H_\rho\approx0.3\,H_{\rho0}$.
Thus, locally we have a wavelength of $\approx10\,H_\rho$.
The situation is similar in the next panel of \Fig{AB}
where the wavelength
is $\approx6\,H_{\rho0}$, and the structures are at a depth
where $H_\rho\approx1.5\,H_{\rho0}$, so locally we have a
wavelength of $\approx9\,H_\rho$.
We can therefore conclude that our earlier results for isothermal
layers can still be applied locally to polytropic layers.

A new aspect, however, that was not yet anticipated at the time,
concerns the importance of NEMPI for vertical fields.
While NEMPI with horizontal magnetic field
still leads to downflows in the nonlinear regime
(the ``potato-sack'' effect),
our present work now confirms that structures consisting of
vertical fields do not sink, but reach a strength comparable to
or in excess of the equipartition value \citep{BKR13,Jab2}.
This makes NEMPI a viable mechanism for spontaneously producing
magnetic spots in the surface layers.
Our present study therefore supports ideas about
a shallow origin for active regions
and sunspots \citep{B05,BKR10,KKWM10,SN12}, contrary to common thinking that
sunspots form near the bottom of the convective zone \citep{Par75,SW80,DSC93}.
More specifically, the studies of \cite{Losada2} point toward the possibility
that magnetic flux concentrations form in the top $6\Mm$, i.e., in the
upper part of the supergranulation layer.

There are obviously many other issues of NEMPI that need to
be understood before it can be applied in a meaningful way
to the formation of active regions and sunspots.
One question is whether the hydrogen ionization layer and
the resulting H$^-$ opacity in the upper layers of the Sun
are important in providing a sharp temperature drop and whether
this would enhance the growth rate of NEMPI, just like strong
density stratification does.
Another important question concerns the relevance of a radiating
surface, which also enhances the density contrast.
Finally, of course, one needs to verify that the assumption of
forced turbulence is useful in representing stellar convection.
Many groups have considered magnetic flux concentrations using
realistic turbulent convection \citep{KKWM10,Rem11,SN12}.
However, only at sufficiently large resolution can one expect
strong enough scale separation between the scale of the smallest
eddies and the size of magnetic structures.
That is why forced turbulence has an advantage over convection.
Ultimately, however, such assumptions should no longer be necessary.
On the other hand, if scale separation is poor, our present
parameterization might no longer be accurate enough, and one would
need to replace the multiplication
between $\qp$ and $\meanBB^2$ in \Eq{dUmean} by a convolution.
This possibility is fairly speculative and requires a separate investigation.
Nevertheless, in spite of these issues, it is important to emphasize
that the qualitative agreement between DNS and MFS is already
surprisingly good.

\begin{acknowledgements}
This work was supported in part by the European Research Council under the
AstroDyn Research Project No.\ 227952,
by the Swedish Research Council under the project grants
621-2011-5076 and 2012-5797 (IRL, AB), by EU COST Action MP0806,
by the European Research Council under the Atmospheric Research Project No.\
227915, and by a grant from the Government of the Russian Federation under
contract No. 11.G34.31.0048 (NK, IR).
We acknowledge the allocation of computing resources provided by the
Swedish National Allocations Committee at the Center for
Parallel Computers at the Royal Institute of Technology in
Stockholm and the National Supercomputer Centers in Link\"oping, the High
Performance Computing Center North in Ume\aa,
and the Nordic High Performance Computing Center in Reykjavik.
\end{acknowledgements}

%r e f

\end{document}